\documentclass{article}
\PassOptionsToPackage{backend=bibtex}{biblatex}
\usepackage{PRIMEarxiv}

\usepackage[utf8]{inputenc} 
\usepackage[T1]{fontenc}    

\usepackage{url}            
\usepackage{booktabs}       
\usepackage{nicefrac}       
\usepackage{microtype}      
\usepackage{lipsum}
\usepackage{fancyhdr}       
\usepackage{graphicx}       

\usepackage{amsfonts}       
\usepackage{amsmath,amssymb,amsfonts}
\usepackage{amsthm}
\usepackage{lmodern}

\usepackage{mathtools}
\usepackage{mathrsfs}

\usepackage{array}
\usepackage{makecell}
\usepackage{multirow}
\usepackage{tabularx}
\usepackage{ragged2e}
\usepackage{longtable}

\usepackage{pdflscape}
\usepackage{rotating}

\usepackage[title]{appendix}

\usepackage{algorithm}
\usepackage{algorithmicx}
\usepackage{algpseudocode}
\usepackage{listings}

\usepackage[table]{xcolor}
\usepackage{textcomp}
\usepackage{manyfoot}
\usepackage{pgffor}
\usepackage{etoolbox}
\usepackage{geometry}
\usepackage{float}

\usepackage{hyperref}
\usepackage{hypcap}
\usepackage{bbm}   

\graphicspath{{plots/}}     

\title{Statistical inference after variable selection in Cox models: A simulation study
}

\author{
  Lena Schemet\\
  Mathematical Statistics and Artificial Intelligence in Medicine, University of Augsburg\\
  Augsburg, Germany\\
  \texttt{lena.schemet@uni-a.de}
  \And
  Sarah Friedrich-Welz\\
  Mathematical Statistics and Artificial Intelligence in Medicine, University of Augsburg\\
  Center for Advanced Analytics and Predictive Sciences (CAAPS), University of Augsburg\\
  Augsburg, Germany\\
  \texttt{sarah.friedrich@uni-a.de}
}

\begin{document}
\maketitle

\begin{abstract}
    Choosing relevant predictors is central to the analysis of biomedical time-to-event data. Classical frequentist inference, however, presumes that the set of covariates is fixed in advance and does not account for data-driven variable selection. As a consequence, naive post-selection inference may be biased and misleading. In right-censored survival settings, these issues may be further exacerbated by the additional uncertainty induced by censoring. We investigate several inference procedures applied after variable selection for the coefficients of the Lasso and its extension, the adaptive Lasso, in the context of the Cox model.  The methods considered include sample splitting, exact post-selection inference, and the debiased Lasso. Their performance is examined in a neutral simulation study reflecting realistic covariate structures and censoring rates commonly encountered in biomedical applications.  To complement the simulation results, we illustrate the practical behavior of these procedures in an applied example using a publicly available survival dataset.
\end{abstract}

\keywords{Survival analysis; Cox model; Lasso; Inference after variable selection; Post-selection Inference; Debiased Lasso}






\section{Introduction}\label{sec::intro}

Variable selection has become an integral component of modern regression analysis,
particularly in biomedical research where complex covariate structures and limited
sample sizes are common.
Methods such as the Lasso \cite{tibshirani1996regression} and its variants
\cite{zou2006adaptive} are widely used to identify relevant predictors while performing
regularization to stabilize estimation.
However, classical inferential procedures are not designed to accommodate the
data-driven nature of variable selection.
As a consequence, naive confidence intervals and $p$-values computed after model
selection may be severely biased, anticonservative, and ultimately misleading
\cite{Berk2013PostSelection}.

This problem is well known in linear and generalized linear models, and a growing body
of work has highlighted the need for inference procedures that explicitly acknowledge
the selection step \cite{lee2016exact}.
In the context of time-to-event outcomes the issue is even more pronounced.
Right censoring introduces additional uncertainty, and the dependence of event times on
both covariates and censoring mechanisms complicates the use of standard asymptotic
arguments \cite{andersen1993model}.
Despite the widespread use of penalized Cox models in biomedical applications
\cite{tibshirani1997lasso, zhang2007adaptive} principled inference for their coefficients
after variable selection remains challenging and comparatively underexplored.


Several methodological proposals have been developed to address these limitations.
Sample splitting offers a conceptual toy strategy by separating model selection
from subsequent inference.
Its effectiveness typically relies on sufficiently large sample sizes, since both steps
are carried out on only a subset of the available data which may reduce statistical
efficiency \cite{cox1975note}.
Post-selection inference conditions explicitly on the selected model and can yield valid
post-selection confidence intervals in regression settings.
General frameworks for valid and exact post-selection inference have been developed and
theoretically justified primarily for linear regression models
\cite{Berk2013PostSelection, lee2016exact, taylor2015statistical}.
Extensions of these approaches to the Cox model are currently
limited \cite{yu2021coxci, kong2021robustcox}.
In particular, exact post-selection inference procedures for Cox regression have so far
been studied only under fixed shrinkage and without randomization.
Randomized post-selection inference has been shown to improve power and stability in
linear models \cite{fithian2014optimal, taylor2018post}, but corresponding theory and
implementations are not yet established for the Cox model.

In addition to selective inference methods that explicitly condition on the selection event, there exist debiasing approaches for inference after variable selection.
These methods apply approximate one-step corrections to penalized estimators in order to restore asymptotic normality.
As a result, inference can be carried out after variable selection without explicitly conditioning on the selected model. Their theoretical properties and empirical performance have been predominantly
investigated for linear regression models \cite{zhang2014confidence, vandegeer2014optimal}.
Consequently, despite substantial progress in low- and high-dimensional linear models,
the available evidence for the validity and practical performance of inference after
variable selection in survival analysis remains comparatively sparse
\cite{yu2021coxci, kong2021robustcox}.

Motivated by these considerations, we systematically evaluate inference procedures
applied after variable selection for Lasso-type estimators in Cox models.
Building on the framework of Kammer et al.\ \cite{Kammer2022LassoSelective}, who assessed
inference after variable selection in Gaussian settings, we extend their simulation
design to right-censored survival data and investigate how censoring, covariate
dependence, and model sparsity affect the validity and efficiency of competing
approaches.
The methods examined include sample splitting, exact post-selection inference, and the
debiased Lasso.

Our main contributions are:
\begin{itemize}
    \item We adapt and extend the simulation study of Kammer et al.\ \cite{Kammer2022LassoSelective} to the Cox model, incorporating realistic survival and censoring structures relevant to biomedical applications.
    \item We provide a comparative evaluation of several inference-after-selection approaches, focusing on coverage, confidence interval width, and selective power under a wide range of scenarios.
    \item We illustrate the practical implications of these methods in an applied example using publicly available survival data.
\end{itemize}

The remainder of the paper is organized as follows.
Section~\ref{sec::methods} introduces the Cox model setting, the Lasso and adaptive Lasso estimators, and
the inference procedures considered, including sample splitting, selective inference,
and debiasing approaches.
Section~\ref{sec::simstudy} describes the simulation study design, including data-generating mechanisms,
estimands, and performance measures.
Section~\ref{sec:results} presents the simulation results.
Section~\ref{sec:example} illustrates the proposed methods using a real data example.
Section~\ref{sec:discussion} concludes with a discussion of the main findings and implications.

Overall, our results offer guidance for practitioners seeking reliable inference after
Lasso-based variable selection in survival analysis and highlight methodological aspects
that warrant further research.

\section{Methods}\label{sec::methods}

This section describes the methodological framework of the study.
We first introduce the model setting and notation after which we present the variable
selection procedures and selective inference methods considered.

\subsection{Setting}
Let $X = (X^{(1)}, \ldots, X^{(p)})^\top$ denote a $p$-dimensional covariate vector, with $(\cdot)^\top$ denoting the transpose of any vector of interest, and let $T$ denote the failure time of interest.

Under right censoring with censoring time $C$, we observe the survival time
$Y = \min(T, C)$
and the event indicator
$\delta = \mathbbm{1}(T \le C)$.
We assume non-informative censoring in the sense of Andersen et al.~\cite{andersen1993model},
that is, the censoring mechanism is independent of the event process conditional on the
observed covariates and does not depend on the unknown regression parameters.

For $n$ independent individuals, we observe i.i.d.\ data
$(Y_i, \delta_i, X_i)$ for $i = 1, \ldots, n$, where
$X_i = (X_i^{(1)}, \ldots, X_i^{(p)})^\top$ denotes the $p$-dimensional covariate vector.

The Cox model \cite{cox1972regression} specifies the
conditional hazard at time $t$ as
\begin{equation*}
    h(t \mid X) = h_0(t) \exp(X^\top \beta^0),
\end{equation*}
where $h_0(t)$ is the unknown baseline hazard function and
$\beta^0 = (\beta_1^0, \ldots, \beta_p^0)^\top$ denotes the unknown true vector of
regression coefficients.

Given the observed data the log partial likelihood, viewed as a function of a 
parameter $\beta \in \mathbb{R}^p$, is given by
\begin{equation}
\label{eq:partiallikeli}
    \ell(\beta)
    = \sum_{i=1}^n \delta_i \left(
        X_i^\top \beta
        - \log \sum_{j : Y_j \ge Y_i} \exp(X_j^\top \beta)
      \right).
\end{equation}

Estimation of the true regression parameter $\beta^0$ is based on maximizing the log
partial likelihood \eqref{eq:partiallikeli}, yielding an estimator $\hat{\beta}$.
In settings with many covariates, penalized approaches such as the Lasso and its variants
have become standard tools \cite{tibshirani1997lasso, fan2002variable, zhang2007adaptive}.

\subsection{Lasso}

The Lasso estimator \cite{tibshirani1996regression}, adapted to the Cox model, is obtained
by maximizing the $\ell_1$-penalized partial likelihood,
\begin{equation*}
    L_P(\beta) = \ell(\beta) - \lambda \sum_{j=1}^p |\beta_j|,
\end{equation*}
where $\lambda > 0$ is the regularization parameter.

The Lasso has several attractive properties \cite{zhang2007adaptive}.
In particular, the $\ell_1$-penalty performs variable selection by shrinking some
coefficients exactly to zero.
The method has become widely used in linear regression and increasingly in Cox models,
especially in biomedical applications involving large-scale covariate data \cite{tang2017spikeslab,wang2025lassocox}.

However, these advantages come with well-known limitations:
\begin{enumerate}
    \item The Lasso tends to include too many variables, leading to an inflated
    false-positive rate in variable selection \cite{Wainwright2009Sharp}.
    \item Estimated coefficients are biased toward zero, especially for variables with
    larger true effects \cite{zhang2014confidence, hastie2015statistical}.
\end{enumerate}

One approach to mitigate this shrinkage bias is the adaptive Lasso \cite{zou2006adaptive},
which incorporates data-dependent weights $w_j$, $j \in \{1,\ldots,p\}$, into the penalty
term.
A common choice is to define these weights as inverse functions of preliminary coefficient
estimates, for example $w_j = 1 / |\hat{\beta}_j|^\gamma$ with $\gamma \in \mathbb{N}$.
This weighting scheme reduces shrinkage for larger coefficients while preserving sparsity
among smaller ones.
Under this choice, the adaptive Lasso estimator achieves $\sqrt{n}$-consistency and
satisfies the oracle property \cite{zhang2007adaptive}.

Despite these improvements, adaptive Lasso procedures do not always achieve satisfactory
coverage properties in inference after variable selection for linear regression
\cite{Berk2013PostSelection}.
Motivated by this, we consider and compare both the standard Lasso and the adaptive Lasso
in the Cox model setting.

\subsection{Conceptual framework: selective inference after variable selection}

Selective inference addresses inferential questions that arise when hypotheses,
models, or inferential targets are not specified prior to observing the data, but are
generated as part of the data-analytic process itself
\cite{Berk2013PostSelection, taylor2015statistical}.
This setting naturally arises in variable selection problems, where the data are first
used to choose a subset of covariates and inference is subsequently reported only for
the selected variables.

In the context of variable selection, the data are used to select a submodel
$M \subseteq M_F$, where $M_F$ denotes the full set of candidate predictors, for example
via the Lasso or adaptive Lasso.
Inference is then performed for regression coefficients associated with the selected
variables.
Classical frequentist inference, however, constructs $(1-\alpha)$ confidence intervals
for full-model parameters $\beta_j^F$ such that
\[
\mathbb{P}\!\left( \beta_j^F \in CI_j \right) = 1-\alpha,
\]
under the assumption that the inferential procedure is independent of any data-driven
model choice.
Once variable selection is introduced, these guarantees generally fail, because the
selection step alters the distribution of estimators and test statistics
\cite{LeebPotscher2005, LeebPotscher2006}.

\subsubsection{From the full-model world to the submodel world}
Variable selection fundamentally changes the inferential perspective.
After a data-driven selection step, inference is no longer conducted in the full-model
parameter space, but in a submodel world that is induced by the selection procedure.
In this submodel world, regression coefficients are defined only relative to the
selected set of variables and generally differ from their full-model counterparts.

For a given submodel $M \subseteq M_F$, the associated population-level coefficient
vector is
\[
\beta_{M}
=
(\beta_{j,M})_{j \in M}
=
\beta_M(\beta^F),
\]
which can be viewed as the projection of the full-data parameter $\beta^F$ onto the
submodel corresponding to $M$
\cite{fithian2014optimal, lee2016exact}.
In linear regression, these submodel coefficients admit closed-form expressions,
whereas in the Cox model no such analytic representation is available
\cite{andersen1993model, therneau2000cox}.
Instead, they are implicitly defined through the partial likelihood and obtained
numerically, for example via Newton--Raphson–type algorithms
\cite{cox1972regression, andersen1993model}.

\subsubsection{Selective confidence intervals as primary estimands}
In this work, we do not take these submodel coefficients themselves as the primary
inferential targets.
Rather, our focus is on the uncertainty statements that are reported after variable
selection. 

We use the term \emph{selective confidence interval} (SCI) to denote an confidence interval 
obtained after a data-driven variable selection step and reported for a selected
variable. Throughout this paper, these SCIs constitute the primary estimands.
Different inference procedures are evaluated and compared through the SCIs they produce,
even though these intervals may rely on different conditioning principles and may be
associated with different underlying reference coefficients.
This operational perspective allows us to compare methods as practical tools for
quantifying uncertainty after selection, despite differences in their formal inferential
targets.

For methods based on sample splitting and exact conditional post-selection inference,
SCIs are constructed conditional on the selected model $\widehat{M}$ and target the
corresponding submodel-specific coefficients $\beta_{j,\widehat{M}}$.
For these approaches, selective coverage is defined conditionally on the selection
event and takes the form
\begin{equation}\label{eq:cond}
\mathbb{P}\!\left(
\beta_{j,\widehat{M}} \in SCI_{j,\widehat{M}}
\;\middle|\;
\widehat{M}
\right)
=
1-\alpha,
\qquad j \in \widehat{M}.
\end{equation}

In contrast, the debiased Lasso constructs SCIs for the full-model coefficients
$\beta_j^0$ without conditioning on the selection event.
As a consequence, its selective coverage is unconditional and asymptotic in nature and
does not correspond to conditional submodel inference.

Although the inference procedures considered in this study differ in their underlying
estimands and coverage guarantees, they address a common practical question:
how to quantify uncertainty for regression coefficients that are selected by a
data-driven variable selection procedure.
Throughout this paper, we therefore adopt a broad, operational definition of
\emph{selective inference} as inference conducted after variable selection and reported
for the selected variables, with SCIs serving as the central objects of comparison.

\subsection{Inference procedures}

We now describe the specific inference procedures considered in this study.
All methods are applied after Lasso variable selection and are used to construct
selective confidence intervals, but they differ in their inferential
targets, conditioning strategies, and coverage guarantees.

\subsubsection{Sample splitting}

The separation of model selection and statistical inference was first discussed by Cox
\cite{cox1975note}.
Sample splitting translates this idea into a practical procedure by dividing the data
into two disjoint parts.

In the first step, one part of the data is used to select a submodel $\widehat{M}$.
In the second step, the remaining data are used to estimate the corresponding submodel
coefficients and to construct SCIs.
Since inference is performed using data that was not involved in the selection step,
the resulting SCIs satisfy the conditional selective coverage property \eqref{eq:cond}.
SCIs are obtained by fitting an unpenalized Cox model to the selected variables using
the inference subsample and constructing standard Wald-type intervals based on the
partial likelihood.

In applied work, a simple 50/50 split of the data is frequently used as a default choice.
This practice is supported by simulation studies in linear regression settings, which
suggest that such a split can provide a reasonable compromise between stable model
selection and efficient inference \cite{fithian2014optimal, tian2018selective}.
However, no comparable guidance is currently available for the Cox model, where the
effective information content depends on the censoring mechanism and the sample size
\cite{yu2021coxci, kong2021robustcox}.
By choosing the split proportion, one can balance selection accuracy and inferential
precision.
For simplicity, we use a 50/50 split throughout.

From an implementation perspective, sample splitting is straightforward and
computationally feasible.
It does not require constrained optimization or matrix inversions beyond those
encountered in standard unpenalized Cox model fitting.

\subsubsection{Exact conditional post-selection inference}

The exact conditional post-selection inference framework (exact PSI) introduced by
Lee et al.\ \cite{lee2016exact} provides finite-sample valid inference after Lasso
selection by conditioning on the selection event.

Inference is carried out conditional on the selected submodel and targets the
corresponding submodel-specific estimands.
As a consequence, the resulting SCIs satisfy the conditional selective coverage
property \eqref{eq:cond}.

For the Cox model adaptations are required due to the structure of
the partial likelihood, but the underlying conditional inference principle remains
conceptually unchanged.
Available software implementations allow exact conditional selective inference
when the penalty parameter $\lambda$ is treated as fixed.

Accordingly, exact PSI is applicable when the regularization parameter is specified
in advance rather than chosen in a data-dependent or prediction-oriented manner.
For linear regression models, randomized extensions of exact PSI have been proposed to
accommodate data-driven tuning parameter selection.
Such randomized procedures are currently unavailable for the Cox model and are
computationally demanding even in the linear setting \cite{Kammer2022LassoSelective}.

\subsubsection{Debiased Lasso}

The debiased Lasso was originally proposed for linear regression models to enable valid
statistical inference by correcting the shrinkage bias induced by $\ell_1$ penalization
\cite{zhang2014confidence, vandegeer2014optimal}.
The approach was subsequently extended to the Cox proportional hazards model by
Lu and Xia \cite{xia2023coxdiverging}.

The central idea is to augment the Lasso estimator with a one-step correction that
removes the leading bias term and thereby restores asymptotic normality.
For the Cox model, this is achieved by replacing the least-squares score function and
information matrix with their counterparts derived from the partial likelihood.
Inference is then based on the score function of the Cox partial likelihood and an approximation of the inverse
Fisher information matrix that accounts for censoring.

Specifically, the debiased estimator for coefficient $j$ is given by
\[
\tilde{\beta}_j
=
\widehat{\beta}_j
+
\widehat{\Theta}_j^\top U(\widehat{\beta}),
\]
where $\widehat{\beta}_j$ is the $j$th component of the Lasso estimator, $U(\widehat{\beta})$ denotes the score function of the Cox partial likelihood evaluated at $\widehat{\beta}$, and $\widehat{\Theta}_j$ denotes
the $j$th row of an estimator of the inverse Fisher information matrix
\cite{xia2023coxdiverging}.

Under suitable regularity conditions, the debiased estimator $\tilde{\beta}_j$ is
asymptotically normal, which enables the construction of SCIs of the form
\[
SCI^{\mathrm{DB}}_j
=
\tilde{\beta}_j
\pm
z_{1-\alpha/2}\,
\widehat{\sigma}_j / \sqrt{n},
\]
where $z_{1-\alpha/2}$ denotes the $(1-\alpha/2)$-quantile of the standard normal
distribution and $\widehat{\sigma}_j$ is a consistent estimator of the asymptotic
standard deviation of $\tilde{\beta}_j$.
For the Cox model, $\widehat{\sigma}_j$ is obtained from an approximation of the inverse
Fisher information matrix $\widehat{\Theta}_j$ based on nodewise regression, which accounts for censoring and
avoids direct inversion in high dimensions \cite{vandegeer2014optimal, xia2023coxdiverging}.

Unlike sample splitting and exact PSI, the debiased Lasso targets the full-model
estimand $\beta_j^F$ (or its partial-likelihood projection) and does not condition on the
selection event.
It therefore performs inference after variable selection without explicitly accounting
for the selection step.

From a computational perspective, the debiased Lasso is more demanding than sample
splitting and exact PSI, as it requires fitting multiple auxiliary regression models to
approximate the inverse Fisher information matrix \cite{vandegeer2014optimal}.
An \textsf{R} implementation for debiased inference in the Cox model is available as
research code by Lu and Xia \cite{xia2023coxdiverging}.

\section{Simulation study}\label{sec::simstudy}

We conducted a comprehensive simulation study to compare selective inference procedures applied
after variable selection in Cox models.
The study is designed to systematically assess how different selective inference
approaches perform under varying sample sizes, covariate dimensions, correlation
structures, censoring levels, and tuning strategies.
Particular emphasis is placed on selective inferential validity, estimation efficiency,
and the relationship between inferential and predictive performance.

The simulation design and reporting follow the ADEMP structure \cite{morris2019using}.

\subsection{Aims (A)}
The aim of this simulation study was to evaluate selective inference procedures for
high-dimensional Cox regression after data-driven variable selection.
We examined the selective coverage of selective confidence intervals (SCIs)
produced by different post-selection inference methods, together with the validity of
the associated selective hypotheses, the correctness and stability of the selected
submodels, and the resulting prediction performance.
All procedures were compared across a broad range of data-generating mechanisms
reflecting typical biomedical survival settings.

\subsection{Data-generating mechanisms (D)}

We conducted a Monte Carlo simulation study with $n_{\text{sim}}=1000$ repetitions per
scenario.
We denote the set of simulation iterations by
$S=\{1,\ldots,n_{\text{sim}}\}$.
Sample sizes $n$ ranged from 75 to 800, starting at $n=75$ and increasing in steps of
100 thereafter (Table~\ref{tab:design_factors}).

\subsubsection{Toy simulation}

For each individual, we generated a $p$-dimensional covariate vector with
$p \in \{10, 20, 50\}$.
Covariates were drawn from a multivariate normal distribution with zero mean and
correlation structure
\[
\Sigma_{ij} = \rho^{|i-j|},\qquad \rho \in \{0.0, 0.3\},
\]
yielding either independent or moderately correlated predictors.
In scenarios designed to mimic mixed data types, a subset of covariates was dichotomized
to obtain approximately Bernoulli$(0.5)$ variables.

The true coefficient vector $\beta^0$ was generated according to one of four
predefined patterns (Table~\ref{tab:beta_patterns}): An all-ones pattern with equal
non-zero effects across all covariates, a high-contrast pattern featuring alternating
large and small effects in the first four coefficients, a realistic pattern with
moderate effects on a limited number of covariates, and a sparse pattern with only one
or two active coefficients.
All remaining coefficients were set to zero, depending on the dimension $p$.

\begin{table}[htbp]
  \centering
  \caption{Coefficient patterns used for the data-generating vector $\beta^0$.}
  \label{tab:beta_patterns}
  \begin{tabularx}{0.9\textwidth}{@{} l l X @{}}
    \toprule
    \textbf{} & \textbf{Coefficient pattern} & \textbf{Description} \\
    \midrule
    allones       & $(1, 1, \ldots, 1)$              & Equal effects on all covariates. \\
    highcontrast  & $(0.3, 1.0, 0.3, 1.0, 0,\ldots,0)$ & Alternating large and small effects. \\
    realistic     & $(0.8, 0.7, 0.5, 0.8, 0,\ldots,0)$ & Moderate effects on a few covariates. \\
    sparse        & $(1, 1, 0,\ldots,0)$              & Two active coefficients. \\
    \bottomrule
  \end{tabularx}
\end{table}

Event times were generated from a Cox model with linear predictor $X^\top\beta^0$ based
on Ramos et al.\ \cite{ramos2020sampling}.
Two baseline hazard distributions were considered.
For the exponential baseline,
\[
T = -\frac{\log U}{\exp(X^\top\beta^0)},\qquad U \sim \mathrm{Unif}(0,1),
\]
which corresponds to a constant baseline hazard.
For the Weibull baseline, survival times were generated by inverse transform sampling as
\[
T=\left(\frac{-\log U}{s\,\exp(X^\top\beta^0)}\right)^{1/k},\qquad
U\sim\mathrm{Unif}(0,1),
\]
which corresponds to a Cox model with cumulative baseline hazard
$H_0(t)=s\,t^{k}$ and baseline hazard $h_0(t)=s\,k\,t^{k-1}$.
Here, $k>0$ denotes the Weibull shape parameter, yielding an increasing baseline hazard
for $k>1$ and a decreasing baseline hazard for $k<1$, while $s>0$ is a scale parameter
controlling the overall timescale of the baseline hazard.

In all Weibull scenarios, we fixed the scale parameter to $s=1$ to simplify the
simulation design.
Since the baseline scale is not identifiable in the Cox model and does not affect the
partial likelihood for the regression coefficients, this choice is not expected to
influence inference on $\beta$ and allows us to focus on selective inference properties
\cite{andersen1982cox, therneau2000cox}.
Moreover, the shape parameter was fixed to $k=2$, corresponding to a moderately
increasing baseline hazard.
Varying the baseline hazard shape was not a primary focus of this study, as inference
after variable selection in Cox models is expected to primarily depend on the regression
structure rather than on the specific parametric form of the baseline hazard.

Independent censoring was imposed through a simple administrative censoring mechanism.
For each replicate, we first generated event times $\{T_i\}_{i=1}^n$ and then defined a
censoring cutoff $c^\star$ as the $(1-\pi_C)$-quantile of $\{T_i\}_{i=1}^n$, where
$\pi_C \in \{0,0.1,0.3\}$ denotes the target censoring proportion.
Observed times were defined as $Y=\min(T,c^\star)$ with event indicator
$\delta=\mathbf{1}(T \le c^\star)$.
This construction yields censoring proportions close to the target level by design \cite{ramos2020sampling}.

To avoid ties, small deterministic jitters were added to duplicated event
(and, where applicable, censoring) times only.
Specifically, for each group of identical times, we added offsets of order
$\varepsilon_T=10^{-8}$ to the event times and $\varepsilon_C=5\cdot 10^{-9}$ to the censoring times,
scaled by the range of the corresponding time variable.

\subsubsection{Realistic simulation}

In addition to the toy settings, we considered a realistic simulation scenario
calibrated to the Molecular Taxonomy of Breast Cancer International Consortium
(METABRIC) study \cite{curtis2012genomic}, a large breast cancer cohort with long-term clinical follow-up.
The same cohort is also used for the real-data illustration in
Section~\ref{sec:example}.
The aim of this design is to retain full control over the data-generating mechanism
while incorporating empirically motivated covariate distributions, effect sizes, and dependence structures.
This approach follows the principles of realistic parametric simulation designs
discussed by Sauer et al.\ \cite{sauer2025parametric}.

We constructed a fixed clinical covariate set reflecting the variables used in the real
data analysis in Section~\ref{sec:example}, including age at diagnosis, tumor stage, ER/HER2 status, progesterone receptor status,
chemotherapy, hormone therapy, radiotherapy, histologic grade, tumor size, number of
positive lymph nodes, and the Nottingham Prognostic Index.
Categorical predictors were encoded via an explicit dummy design matrix without
intercept.
To obtain a stable and reproducible design across simulation replicates, constant
columns and very rare $0/1$ dummy variables were removed, and all remaining covariates
were centered and scaled.
The resulting stabilized design matrix was kept fixed throughout the simulation study.

Event times were generated from a Cox model with a Weibull baseline hazard using inverse
transform sampling, following the same data-generating mechanism as in the toy
simulation.
The Weibull baseline was chosen because it provides a flexible monotone hazard function
and, in preliminary experiments comparing several parametric baseline distributions (including exponential, log-normal, and log-logistic),
yielded the best discrimination performance in terms of time-dependent AUC.
Details of this model comparison are provided in the Supplementary Material~S1.1.

A Weibull model was fitted once to the stabilized METABRIC design matrix, and the fitted
regression coefficients were treated as the simulation truth.
In each simulation replicate, covariate profiles were resampled with replacement from
the METABRIC cohort, and event times were generated from the calibrated Weibull baseline
model.
Target censoring proportions were imposed using the same administrative censoring
mechanism as in the toy simulation, with censoring levels varied across scenarios.

Overall, the METABRIC simulation represents a structured extension of the toy design:
the survival time generation, baseline hazard formulation, and censoring mechanism are
identical, while the covariate distribution and regression coefficients are grounded in
real clinical data.
This allows us to study selective inference procedures under controlled yet
realistically calibrated conditions.

\subsubsection{Design factors}

The simulation grid was defined by the design factors listed in
Table~\ref{tab:design_factors}.
In the toy simulations, each scenario was evaluated under two baseline hazard
distributions (exponential and Weibull) and four coefficient patterns
(all-ones, high-contrast, realistic, and sparse).
In contrast, the METABRIC simulation is based on a single Weibull baseline hazard and
an empirically calibrated coefficient pattern derived from the METABRIC data.
All remaining design factors, including sample size and target censoring proportion,
were varied analogously to the toy settings.

\begin{table}[htbp]
\centering
\caption{Design factors varied in the simulation study.}
\label{tab:design_factors}

\small
\setlength{\tabcolsep}{8pt}
\renewcommand{\arraystretch}{1.2}

\begin{tabularx}{\textwidth}{l X X}
\toprule
\textbf{Factor}
& \textbf{Toy simulations}
& \textbf{Realistic simulations (METABRIC)} \\
\midrule

Sample size $n$
& 75--800  & 75--800 \\

Number of covariates $p$
& $10,\ 20,\ 50$
& $10$ \\

Correlation $\rho$
& $0.0,\ 0.3$
& \textemdash \\

Target censoring proportion
& $0,\ 0.10,\ 0.30$
& $0,\ 0.10,\ 0.30$ \\

Baseline distribution
& Exponential, Weibull
& Weibull \\

Coefficient pattern
& four different settings (Table~\ref{tab:beta_patterns})
& real-data--based pattern \\
\bottomrule
\end{tabularx}
\end{table}

\subsection{Estimands (E)}

The primary inferential target of the simulation study is the collection of
\emph{selective confidence intervals} (SCIs) reported for regression coefficients that
are selected by a data-driven variable selection procedure.
For a selected submodel $\widehat{M}\subseteq M_F$ and a coefficient index
$j\in\widehat{M}$, we denote the corresponding SCI by $SCI_{j,\widehat{M}}$.
In the simulation study, we write $\widehat{M}_s$ for the selected submodel in
iteration $s\in S$ and $SCI_{j,\widehat{M}_s}$ for the SCI reported for variable $j$
in that iteration, provided $j\in\widehat{M}_s$.

Secondary estimands include the associated selective null hypotheses
$H_{0,j}:\beta_j=0$ and the resulting after-selection models.

\subsection{Methods (M)}

All investigated procedures follow a two-stage structure.
In a first step, a subset of covariates is selected using either the standard or the
adaptive Lasso.
In a second step, a dedicated inference procedure is applied to construct selective
confidence intervals (SCIs) for the regression coefficients corresponding to the selected
variables.
Table~\ref{tab:models} provides a structured overview of the considered methods, which
differ in how they account for the variable selection step.

\subsubsection{Baseline and naive comparators}
To establish reference points for comparison, we included a full Cox model containing
all available covariates as well as an oracle model restricted to the truly active
coefficients.

In addition, we considered two naive post-selection refitting strategies that ignore
selection-induced bias.
The first refitting strategy is a standard refit approach, where an unpenalized Cox model is fitted to the
variables selected by the Lasso and inference is based on conventional Wald confidence
intervals obtained from the fully re-optimized partial likelihood.
The second variant, denoted \emph{refit0}, omits the additional Newton--Raphson update
after variable selection and instead relies on the one-step estimator produced at the
end of the penalized fitting procedure.

The inclusion of the refit0 variant is motivated by methodological considerations.
Inspection of the implementation of exact post-selection inference for the Cox model in
the \texttt{selectiveInference} package \cite{tibshirani2017packageselectiveinference}
reveals that the underlying estimator corresponds to such a one-step refit rather than a
fully re-optimized partial likelihood fit.
Including refit0 therefore allows us to disentangle the effect of the estimator itself
from the additional conditioning step used in exact PSI and to assess how closely a naive
one-step refit aligns with the estimators underlying theoretically justified selective
inference procedures.

\subsubsection{Selective inference methods}
Beyond these baseline and naive approaches, we considered several methods that explicitly
aim to provide valid inference after variable selection.
These include sample splitting, which separates model selection and inference across
independent data splits, the debiased Lasso, which applies a projection-based correction
to mitigate shrinkage bias, and exact PSI, which conditions on
the selection event to achieve exact selective validity.

All Lasso-based procedures rely on a choice of the regularization parameter $\lambda$.
The interaction between tuning strategy and post-selection inference validity is a
central aspect of our empirical investigation.

\subsubsection{Choice of tuning rules}
The performance of both the standard and adaptive Lasso depends critically on the choice
of the regularization parameter $\lambda$.
To provide a comprehensive and practically relevant comparison, five tuning rules were
considered, summarized in Table~\ref{tab:tuning}.
These rules can be grouped into prediction-oriented, model-selection-oriented, and fixed
approaches.

First, two prediction-oriented tuning rules based on 10-fold cross-validation (CV) were
used.
The tuning parameter $\lambda_{\mathrm{CV,min}}$ selects the value of $\lambda$ that
minimizes the cross-validated partial likelihood risk, whereas the
$\lambda_{\mathrm{CV,1SE}}$ rule chooses the largest $\lambda$ whose cross-validated risk
lies within one standard error of the minimum.
The latter favors more frugal models while maintaining comparable predictive
performance \cite[Section~7.10.1]{hastie2009elements}.

Second, two information-theoretic tuning rules were employed.
The AIC-type choice $\lambda_{\mathrm{AIC}}$ is defined by minimizing an
Akaike-type information criterion and primarily targets predictive accuracy, whereas the
BIC-type choice $\lambda_{\mathrm{BIC}}$ imposes a stronger penalty on model complexity and
tends to favor sparser models.
Under suitable regularity conditions, BIC-type criteria are known to lead to consistent
variable selection \cite{akaike1974new, schwarz1978estimating, fan2001variable}.

Finally, a fixed regularization parameter $\lambda_{\mathrm{fix}}$ was included as a
benchmark.
This value was derived from a large-scale external dataset ($N=100{,}000$), where
1{,}000 repeated Lasso fits were performed to obtain a stable, scenario-specific estimate
of an appropriate penalty level.
Technical details on the construction of this fixed tuning parameter and the definition of the AIC and BIC criterias are provided in
Supplementary Material~S1.2.

Taken together, this range of tuning rules reflects commonly used choices in applied
Lasso analyses and allows for a systematic assessment how different selection
strategies interact with subsequent inference and of variable selection properties across
the methods listed in Table~\ref{tab:models}.

\begin{table}[htbp]
\centering
\caption{Overview of methods investigated in this study.}
\label{tab:models}
\small
\setlength{\tabcolsep}{6pt}
\renewcommand{\arraystretch}{1.15}
\begin{tabularx}{\textwidth}{@{} l l l X @{}}
\toprule
Method   & Variable selection & Tuning & Inference \\
\midrule
full     & none  & --        & Wald CI \\
oracle   & none  & --        & Wald CI \\
refit    & lasso & $\lambda$-dependent & Wald CI after refitting the selected model \\
refit0   & lasso & $\lambda$-dependent & Wald CI without refitting \\
split    & lasso & $\lambda$-dependent & Wald CI based on sample splitting \\
debiased & lasso & $\lambda$-dependent & Debiased Wald SCI \\
exact psi & lasso & $\lambda$-dependent & Exact conditional SCI \\
\bottomrule
\end{tabularx}
\end{table}

\begin{table}[htbp]
\centering
\caption{Overview of tuning rules used to select the penalty parameter $\lambda$ in this study.}
\label{tab:tuning}
\small
\setlength{\tabcolsep}{6pt}
\renewcommand{\arraystretch}{1.15}
\begin{tabularx}{\textwidth}{@{} l l X @{}}
\toprule
Tuning rule & Tuning type & Description \\
\midrule
min  & CV-based &
$\lambda$ minimizing cross-validated risk \\
1se  & CV-based &
Largest $\lambda$ within one standard error of the minimum CV risk \\
fix  & fixed &
Fixed penalty parameter $\lambda_{\min}$ obtained from Lasso
fit on a large simulated population dataset \\
aic  & information-based &
$\lambda$ selected by Akaike Information Criterion \\
bic  & information-based &
$\lambda$ selected by Bayesian Information Criterion \\
\bottomrule
\end{tabularx}
\end{table}

\subsection{Performance measures (P)}

Performance is evaluated using \emph{selective coverage}, \emph{SCI width},
\emph{selective power}, and \emph{selective type~I error}.
In addition, runtimes (in seconds) were recorded for each method, and model performance
measures were reported.

Selective coverage is defined as the proportion of SCIs that contain the true target
coefficient among the selected variables and is assessed relative to the nominal
90\% level, consistent with the default settings of our main software package.
Coverage rates closer to the nominal level indicate better adjustment for the variable
selection step.

SCI width corresponds to the average length of the selective confidence intervals and
serves as a measure of estimation precision, with shorter intervals indicating higher
precision.

Selective power quantifies the probability of correctly rejecting a selective null
hypothesis when the corresponding coefficient is truly non-zero, whereas selective
type~I error measures the probability of falsely rejecting the null hypothesis when the
true coefficient equals zero.

Detailed definitions and formulas for all primary performance measures are provided in
Supplementary Material~S1.3.



In addition to inferential performance, we evaluated predictive performance of the
selected models as a secondary outcome.
While the concordance index (C-index) is widely used to assess discrimination in
survival analysis, it has been criticized for limited interpretability and sensitivity
to censoring \cite{Blanche2019Cindex}.
We therefore primarily assess predictive accuracy using the integrated Brier score
(IBS), which aggregates time-dependent squared prediction errors over the follow-up
period and captures both discrimination and calibration. The IBS was computed using inverse probability of censoring weights \cite{gerds2006consistent}.
Lower values of the IBS indicate better predictive performance. C-index is reported for completeness and comparability with prior work.

To quantify the efficiency loss induced by variable selection, predictive performance
was evaluated alongside model size and variable selection accuracy.
An oracle model fitted with knowledge of the true active set serves as a theoretical
benchmark. Corresponding oracle results are reported in supplementary analyses.

\subsection{Software and implementation details}

All simulations and analyses were conducted in \textsf{R} (version 4.4.2).
The \texttt{survival} package \cite{therneau2013rpackagesurvival, therneau2000cox} was used for fitting Cox
models and handling time-to-event data.
Penalized estimation via the Lasso and adaptive Lasso was carried out using the
\texttt{glmnet} package \cite{friedman2021packageglmnet, glmnetold, coxnet}.

Exact conditional post-selection inference was implemented via the
\texttt{selectiveInference} package \cite{tibshirani2017packageselectiveinference}.
While this framework supports the Cox model under fixed penalty conditions, procedures
for randomized or data-driven penalty selection are presently limited to the linear
regression setting.

Methods without established, off-the-shelf software support for the Cox model were
implemented manually based on their methodological descriptions in the literature.
In particular, debiased inference for the Cox model was implemented following Lu and Xia
\cite{xia2023coxdiverging}, using their research code as a reference. All custom implementations and simulation code are provided in the supplementary material.

\section{Results}\label{sec:results}

This section presents the results of the simulation study.
We first report inferential performance for the primary selective estimands and then
summarize non-selective performance measures related to prediction and variable
selection. 

\subsection{Primary estimands}

Results for the primary estimands defined in Section~3 are reported in terms of selective
coverage, SCI width, selective power, and selective type~I error. Qualitative assessments of selective coverage, power, and type I error are based on proximity to the nominal level, robustness across increasing sample sizes, and consistency relative to the oracle procedure. No strict numerical cutoffs are imposed, as the interpretation focuses on systematic patterns rather than isolated values.

\subsubsection{Selective coverage}

Selective coverage is evaluated as defined in Section~3.4.
Coverage probabilities closer to the nominal 90\% level indicate more effective
adjustment for the variable selection step. Selective coverage is described as “high” when it remains close to the nominal 90 \% level with little variation across sample sizes, as “moderate” when noticeable but non-systematic deviations occur, and as “low” when substantial or persistent undercoverage is observed.

\begin{figure}[h]
    \centering
    \includegraphics[width=\linewidth]{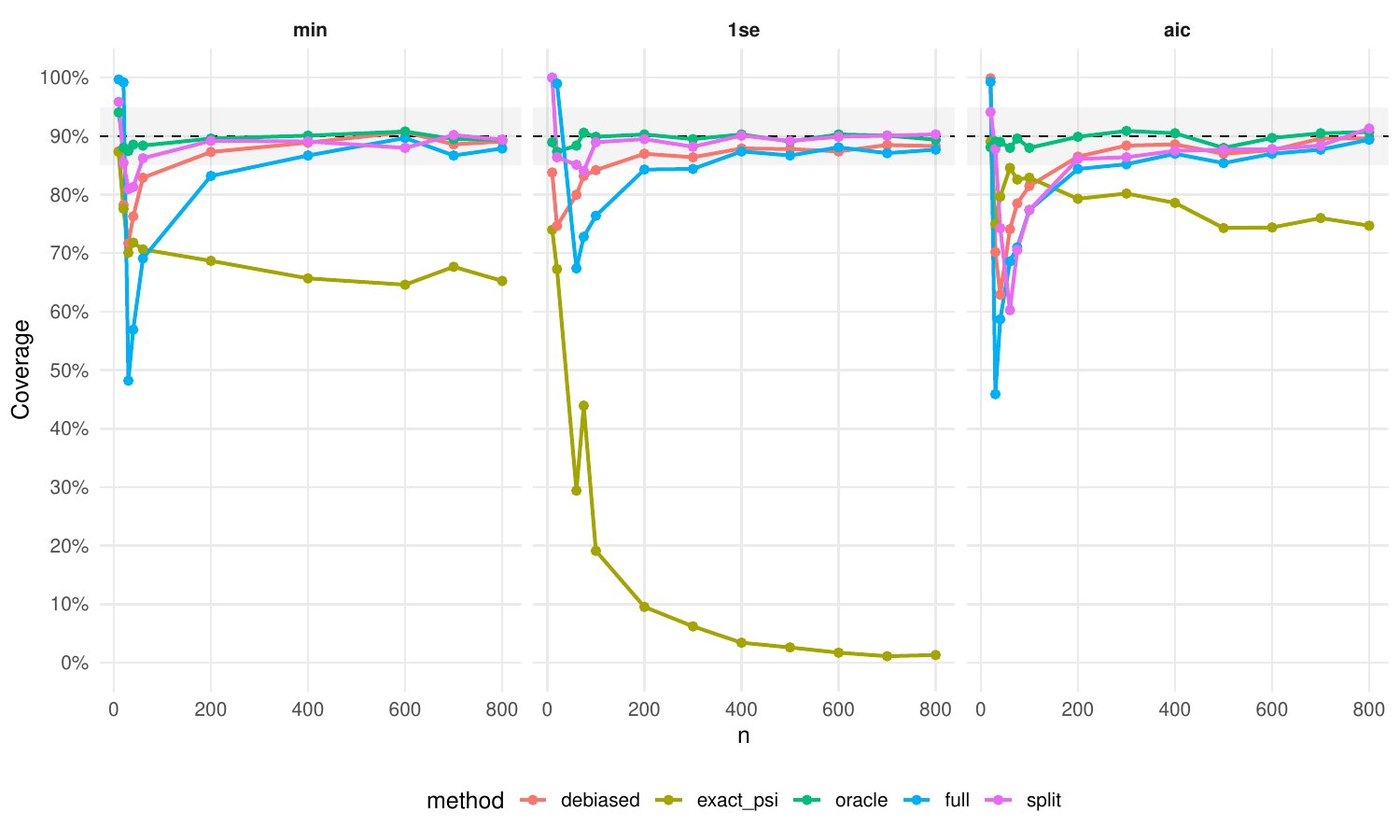}
    \caption{Selective coverage under the realistic coefficient pattern with $p=20$,
    Weibull baseline hazard, no censoring, and correlation $\rho=0.3$.
    Results are shown for coefficient $X_1$ using the non-adaptive Lasso with tuning choices 
    $\lambda_{\mathrm{CV,min}}$, $\lambda_{\mathrm{CV,1se}}$,$\lambda_{\mathrm{AIC}}$.}
    \label{fig:cov_real_p20_x1_rho03}
\end{figure}

Figure~\ref{fig:cov_real_p20_x1_rho03} illustrates selective coverage probabilities for a
representative setting with a realistic coefficient pattern and $p=20$.
Across methods and tuning strategies, selective coverage is generally close to the
nominal level when explicit selection adjustment is applied.

In particular, sample splitting and the debiased Lasso achieve coverage close to the
target level for moderate to large sample sizes ($n \gtrsim 300$) across most simulation
scenarios.
An exception occurs for the $\lambda_{\mathrm{CV,1SE}}$ tuning rule, where increased
variability in model selection leads to modest deviations from nominal coverage.

In contrast, exact PSI exhibits pronounced undercoverage when combined with
cross-validation--based tuning, especially for $\lambda_{\mathrm{CV,min}}$ and
$\lambda_{\mathrm{CV,1SE}}$.
This behavior reflects the sensitivity of exact conditional post-selection inference to
data-driven tuning of the regularization parameter, which violates the fixed-$\lambda$
assumption underlying the method.

These qualitative patterns persist across alternative sample sizes, tuning strategies
($\lambda_{\mathrm{fix}}$, $\lambda_{\mathrm{AIC}}$, $\lambda_{\mathrm{BIC}}$, $\lambda_{\mathrm{CV,min}}$, $\lambda_{\mathrm{CV,1SE}}$), and target
censoring proportions, as documented in the Supplementary Material~S2.1.

\subsubsection{SCI widths}

SCI width is used as a measure of inferential precision, with shorter intervals
indicating greater efficiency.

\begin{figure}[h]
    \centering
    \includegraphics[width=\linewidth]{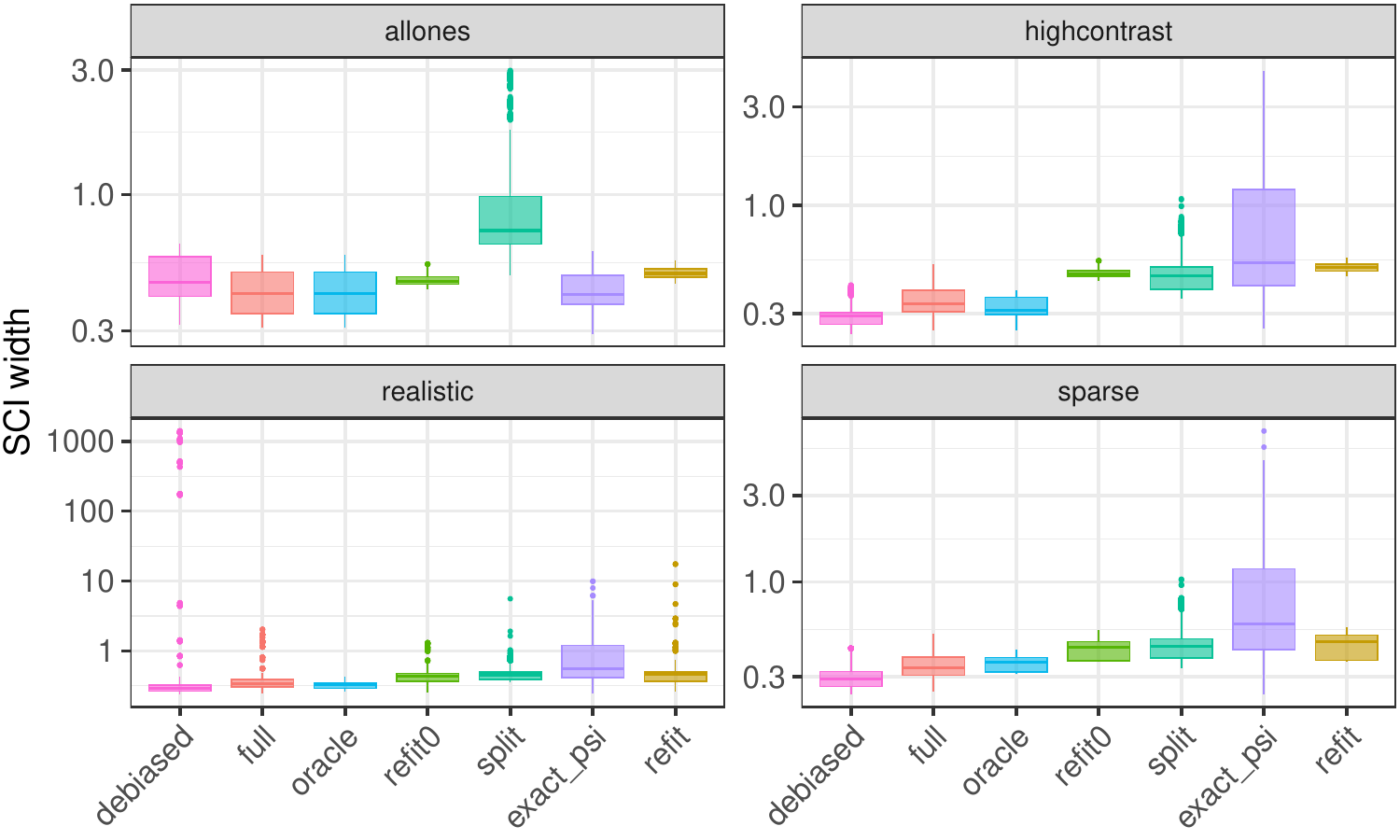}
    \caption{Distribution of SCI lengths for the primary estimand at sample size $n=200$
    across inference methods.
    Results are shown for the toy and METABRIC-calibrated settings on a log scale.}
    \label{fig:ci_width}
\end{figure}

Figure~\ref{fig:ci_width} summarizes the distribution of SCI lengths across inference
methods for a representative toy setting.
SCI width provides a complementary perspective on selective inference performance by
highlighting the practical consequences of different adjustment strategies.

Across scenarios, the debiased Lasso produces comparatively short and stable SCIs,
reflecting a favorable balance between bias correction and variance inflation.
Sample splitting yields wider SCIs with increased variability, particularly under the
METABRIC-calibrated design, consistent with its reduced effective sample size.

Exact PSI produces the widest SCIs overall, especially when combined with
cross-validation--based tuning.
This reflects substantial uncertainty inflation due to conditioning on complex selection
events.
Increasing the sample size uniformly reduces SCI widths for all methods without altering
their relative ranking.

\subsubsection{Selective power and type~I error}

Selective power and selective type~I error are evaluated as defined in Section~3.4 and
quantify the ability to detect non-zero effects while controlling false rejections under
variable selection.

Overall, methods that explicitly account for the selection step achieve a substantially
better balance between selective power and type~I error control than naive post-selection
approaches.
In particular, the debiased Lasso maintains selective type~I error rates close to the nominal level while retaining high power, whereas sample splitting exhibits inflated type~I error rates despite reasonable power.
In contrast, naive refitting and non-selective procedures tend to be liberal.

\begin{figure}[h]
    \centering
    \includegraphics[width=\linewidth]{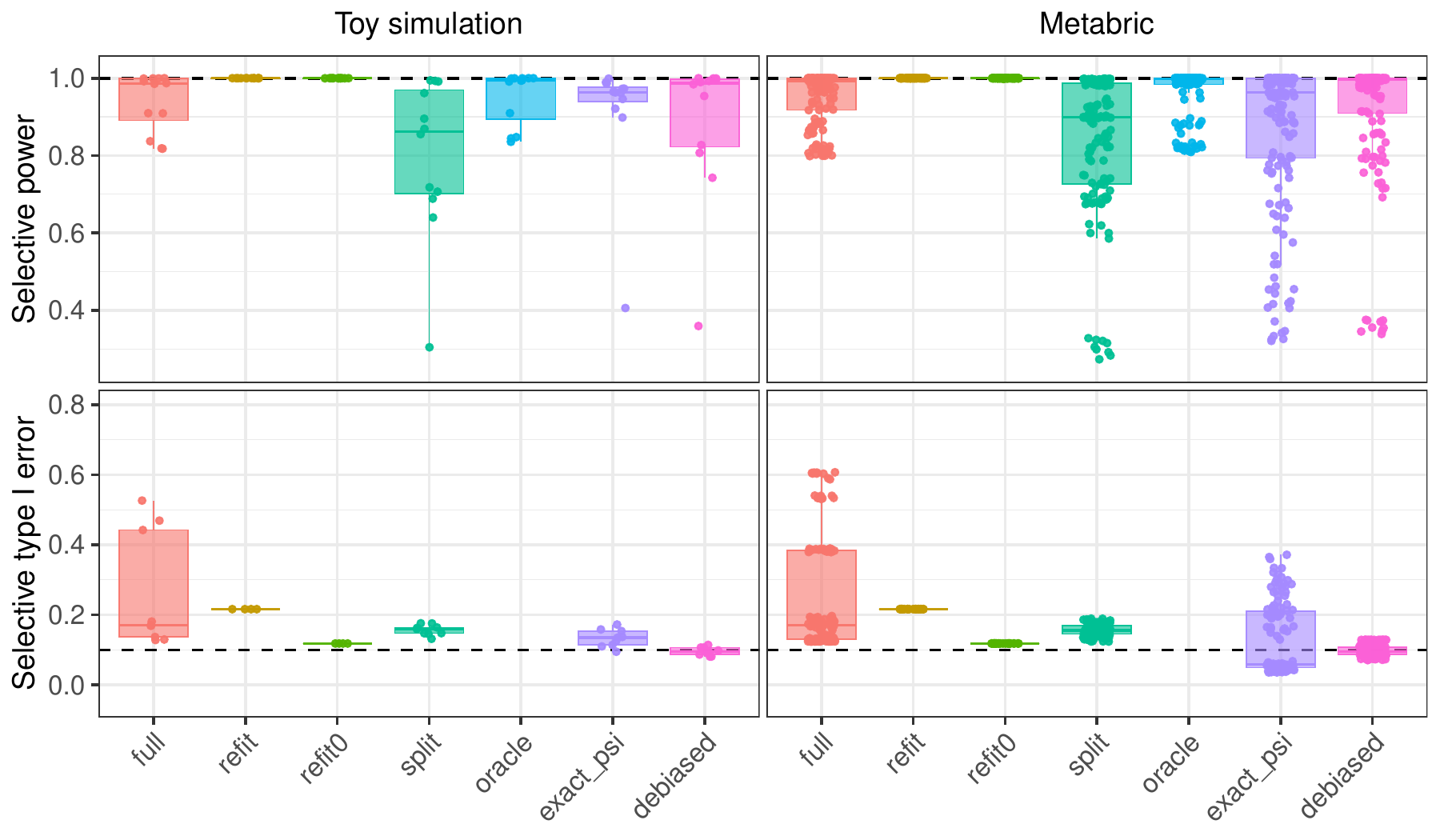}
    \caption{Selective power (top row) and selective type~I error rates (bottom row) for
    the toy and METABRIC settings at sample size $n=75$.
    The dashed horizontal line indicates the nominal type~I error level.}
    \label{fig:power}
\end{figure}

Figure~\ref{fig:power} reports selective power and selective type~I error rates averaged
over multiple simulation scenarios for the toy and METABRIC designs at sample size
$n=75$.
Averaging across scenarios reduces setting-specific variability and highlights systematic
differences between inference procedures.

Results for larger sample sizes ($n=100$ and $n=200$), alternative tuning rules ($\lambda_{\mathrm{CV,min}}$, $\lambda_{\mathrm{CV,1se}}$,
$\lambda_{fix}$, $\lambda_{\mathrm{AIC}}$, $\lambda_{\mathrm{BIC}}$), and different target censoring proportions are reported in the
Supplementary Material~S2.3.
Across this broader range of settings, the qualitative conclusions remain stable.

Increasing the sample size improves selective power and reduces variability in type~I
error rates without changing the relative ranking of the methods.
Less regularized tuning strategies, such as $\lambda_{\mathrm{CV,min}}$ or $\lambda_{\mathrm{fix}}$
rules, yield higher selective power but exhibit greater variability in type~I error
control, particularly under the METABRIC design.
More conservative choices improve stability at the cost of reduced power.
Higher levels of censoring uniformly reduce selective power and amplify method-specific
differences.

\subsection{Performance measures}

Predictive performance of the selected models was evaluated as a secondary outcome.
Results are summarized using the integrated Brier score (IBS), with lower values
indicating better predictive accuracy.
Selection-related metrics such as average model size and the proportion of truly active
variables among the selected set ($P_{\text{true}}$) are reported alongside IBS.
Corresponding results for the concordance index (C-index) are provided in
Supplementary Material~S2.5.

Across simulation settings, prediction-oriented tuning rules, in particular the
AIC-type criterion and $\lambda_{\mathrm{CV,min}}$, most often achieved the lowest IBS.
However, these rules selected substantially larger models and yielded markedly lower
$P_{\text{true}}$, especially for $p=50$.

More parsimonious tuning rules, such as $\lambda_{\mathrm{CV,1SE}}$ and the BIC-type
criterion $\lambda_{\mathrm{BIC}}$, selected considerably smaller models and achieved higher values of
$P_{\text{true}}$, while incurring only a modest loss in predictive performance.
Overall, these results illustrate the expected trade-off between prediction accuracy
and sparsity: tuning strategies optimized for prediction tend to favor dense models,
whereas selection-oriented rules yield more interpretable solutions with improved
selection quality.

\subsection{Summary of main results}

Across all simulation scenarios, inference procedures that explicitly account for the
variable selection step provide substantially improved control of selective coverage and
selective type~I error compared to naive post-selection approaches.
The debiased Lasso and sample splitting achieve coverage close to the nominal level over
a wide range of sample sizes, tuning strategies, and censoring proportions.
In contrast, exact PSI exhibits pronounced sensitivity to data-driven tuning of the
regularization parameter, particularly under cross-validation.

In terms of inferential efficiency, the debiased Lasso consistently yields shorter SCIs
than sample splitting and exact PSI, whereas exact PSI incurs substantial SCI inflation
due to conditioning on complex selection events.
Differences between methods diminish with increasing sample size but remain visible in
moderate-dimensional settings.

Accordingly, type~I error is described as “controlled” when empirical rates remain close to the nominal level, as “inflated” when systematic over-rejection is observed, and as “conservative” when rejection rates are consistently below the nominal level. Selective power and selective type~I error exhibit clear and method-specific trade-offs, which are summarized qualitatively in Table~\ref{tab:primary-estimands}. In this summary, power is described as high or moderate relative to the oracle benchmark, while type~I error behavior is classified as controlled, inflated, or conservative based on systematic deviations from the nominal level.

\rowcolors{1}{white}{gray!12}
\begin{longtable}{l l c c c}
\caption{Qualitative summary of selective inference performance across methods and tuning
strategies.}
\label{tab:primary-estimands}\\
\toprule
Method & Tuning category & Selective coverage & Power & Type~I error \\
\midrule
\endfirsthead
\toprule
Method & Tuning category & Selective coverage & Power & Type~I error \\
\midrule
\endhead
\bottomrule
\endlastfoot
split    & fixed & close to nominal & moderate & inflated \\
         & CV    & close to nominal & moderate & inflated \\
debiased & fixed & close to nominal & high     & controlled \\
         & CV    & close to nominal & high     & controlled \\
exact PSI & fixed & below nominal & low      & conservative \\
          & CV    & below nominal & low      & conservative \\
\end{longtable}

Predictive performance, as measured by the IBS, is comparatively insensitive to
the choice of inference procedure and tuning strategy.
Across methods, predictive accuracy remains relatively stable, even when inferential
properties differ substantially, and efficiency losses relative to the oracle model are
moderate.
Overall, the results highlight a clear distinction between inferential validity and
predictive performance in Cox models after variable selection.

\section{Real data example}\label{sec:example}

We illustrate the proposed selective inference framework using data from the
Molecular Taxonomy of Breast Cancer International Consortium (METABRIC) study
\cite{curtis2012genomic}, a large breast cancer cohort with long-term clinical
follow-up.
Clinical data were obtained from cBioPortal.
Only publicly available clinical variables were used to ensure full reproducibility of
the analysis. No controlled-access molecular or raw genomic data were accessed.

\subsection{Research question}

The METABRIC data example serves as an illustration of variable selection and
post-selection inference in a clinically relevant right-censored survival setting.
The focus of this analysis is methodological rather than etiological: the goal is not
causal interpretation or biomarker discovery, but prognostic modeling based on
routinely available clinical covariates.

Specifically, we aim to investigate:
(i) the stability of variable selection across different Lasso-based methods and tuning
strategies,
(ii) the resulting selective confidence intervals (SCIs) for selected covariates, and
(iii) how different post-selection inference approaches quantify uncertainty in a
data-adaptive modeling workflow.

Overall survival time (in months) was defined as
\texttt{overall\_survival\_months}, and the event indicator was derived from
\texttt{patients\_vital\_status} (death vs.\ censoring).
We considered routinely available clinical predictors, including age at diagnosis,
tumor stage, ER status, HER2 status, PR status, chemotherapy, hormone therapy,
radiotherapy, histologic grade, tumor size, number of positive lymph nodes, and the
Nottingham Prognostic Index (NPI).

\subsection{Analysis}

The METABRIC dataset was analyzed using the same inferential procedures as in the
simulation study, including the full Cox model, sample splitting, the debiased Lasso,
and exact PSI.

To assess the stability of variable selection, we followed an approach similar to
Kammer et al.\ \cite{Kammer2022LassoSelective}.
Specifically, we performed 100 subsampling repetitions.
In each repetition, the variable selection step was re-run, and selection frequencies
were computed as the proportion of subsamples in which each covariate was selected.

In parallel, confidence intervals were recomputed for each subsample and inference
method.
This allows a joint assessment of selection stability and uncertainty quantification
across methods, rather than relying on a single realization of the selected model.

Unless stated otherwise, results are reported for the non-adaptive Lasso with the
cross-validated tuning choice $\lambda_{\mathrm{CV,min}}$, which reflects a commonly
used applied workflow.
Alternative tuning rules ($\lambda_{\mathrm{AIC}}$, $\lambda_{\mathrm{CV,1SE}}$) yield qualitatively
similar patterns and are reported in the Supplementary Material.

\subsection{Results}

\begin{figure}[h]
  \centering
  \includegraphics[width=\linewidth]{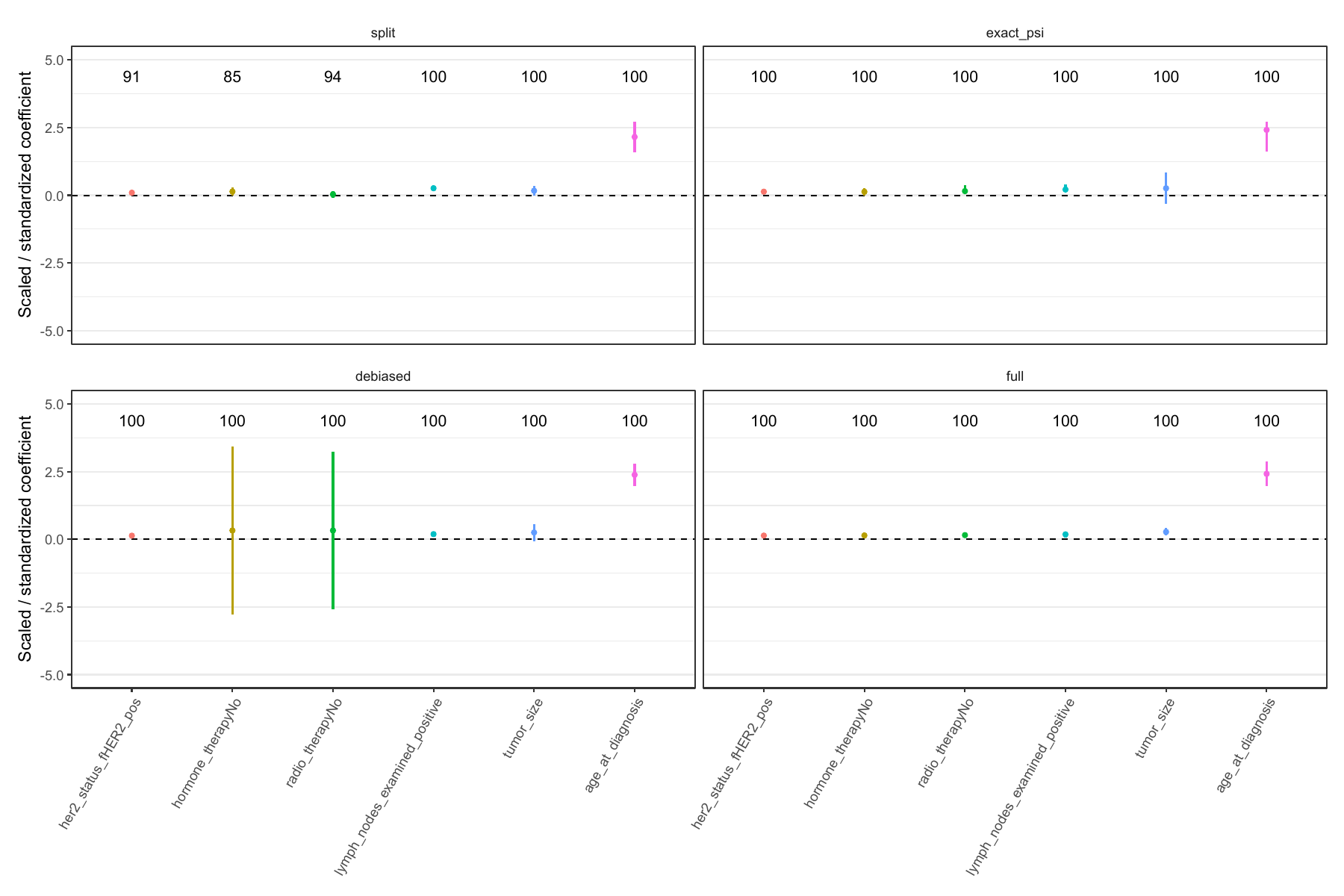}
  \caption{Real data example METABRIC: point estimates and 90\% selective confidence
  intervals for regression coefficients obtained with different inference methods.
  Results are shown for the cross-validated tuning choice $\lambda_{\mathrm{CV,min}}$.
  Coefficients are displayed on the original scale and ordered by increasing standardized
  effect size.
  Numbers above the panels indicate selection frequencies (in \%) across 100 subsamples.}
  \label{fig:realdata_pointest_ci}
\end{figure}

Figure~\ref{fig:realdata_pointest_ci} summarizes the estimated regression coefficients and
corresponding 90\% selective confidence intervals for the METABRIC data.
The presentation follows the layout introduced by Kammer et al.~\cite{Kammer2022LassoSelective}.
Results are shown for the cross-validated tuning choice $\lambda_{\mathrm{CV,min}}$; the
corresponding figures for $\lambda_{\mathrm{CV,1SE}}$ and the AIC-based tuning rule were
qualitatively very similar and are therefore omitted for brevity.

Across inference methods, point estimates are broadly comparable, whereas the width and
stability of the selective confidence intervals differ substantially.
Sample splitting and exact PSI tend to produce wider intervals overall, reflecting the
explicit adjustment for the variable selection step.
In contrast, the debiased Lasso often yields comparatively narrower intervals, albeit with
increased variability across subsamples.

A more detailed inspection reveals that these patterns depend strongly on the covariate
type.
For continuous predictors, the debiased Lasso generally provides more precise selective
inference, resulting in relatively narrow confidence intervals.
However, for binary or ordinal covariates, the debiased Lasso exhibits pronounced interval
inflation, leading to substantially wider confidence intervals.
Exact PSI shows the opposite behavior: while it produces wider intervals for continuous
covariates than the debiased Lasso, it yields more stable and interpretable selective
confidence intervals for binary and ordinal predictors.

Selection frequencies further support this distinction.
Core clinical predictors such as tumor size and tumor stage are selected consistently
across subsamples and inference procedures.
In contrast, treatment-related covariates (e.g., chemotherapy or radiotherapy) exhibit
considerable method-dependent variability, reflecting both correlation structures in the
data and differences in how selection uncertainty is propagated into inference.

Overall, the METABRIC example illustrates how selective inference can be used to
distinguish between stable and unstable prognostic signals in a realistic clinical
setting.
Rather than interpreting individual confidence intervals in isolation, the joint
consideration of selection frequencies and selective confidence intervals provides a
transparent summary of both variable importance and inferential uncertainty in a
data-adaptive modeling pipeline.

\section{Discussion}\label{sec:discussion}

In this work, we conducted a neutral comparison study of selective inference procedures
applied after Lasso-type variable selection in Cox proportional hazards models.
Our objective was not to identify a universally optimal method, but to clarify the
inferential guarantees, efficiency trade-offs, and practical sensitivities that arise
when data-driven variable selection and tuning are combined with post-selection
uncertainty quantification in survival analysis.

A central finding across simulation scenarios is that valid uncertainty quantification
after variable selection critically depends on aligning the inferential target with the
intended analysis goal.
In many applied settings, particularly exploratory or prognostic modeling, the primary
interest lies in the coefficients of the selected submodel, rather than in recovering a
single underlying “true” full model.
From this submodel-oriented perspective, selective inference is appealing because it
explicitly accounts for the optimism introduced by variable screening and supports more
reproducible reporting of uncertainty.

Our simulation results demonstrate that inference procedures which explicitly adjust
for the selection step provide substantially improved control of selective coverage and
selective type~I error compared to naive post-selection refitting.
In particular, the debiased Lasso and sample splitting achieve selective coverage close
to the nominal level over a wide range of sample sizes, censoring levels, and tuning
strategies.
In contrast, exact PSI exhibits pronounced
sensitivity to data-driven tuning choices, especially when the regularization parameter
is selected via cross-validation.
This behavior reflects the fact that exact PSI relies on conditioning on a fixed
selection event, an assumption that is violated when $\lambda$ is chosen adaptively.

At the same time, our results highlight that selection adjustment is not cost-free.
Methods that condition strongly on the selection event tend to produce wider selective
confidence intervals, reflecting substantial uncertainty inflation.
Exact PSI, in particular, yields the widest intervals in most scenarios, whereas the
debiased Lasso achieves a more favorable balance between bias correction and variance,
resulting in comparatively short and stable intervals.
Sample splitting, while conceptually simple and robust, suffers from a reduced effective
sample size and consequently wider intervals, especially in moderate-to-high censoring
settings.

The choice of tuning strategy for the regularization parameter $\lambda$ emerged as a
key driver of practical performance.
Prediction-oriented rules such as $\lambda_{\min}$ or AIC-type criteria $\lambda_{\mathrm{AIC}}$ typically select
larger models, leading to higher selective power but increased variability in selective
type~I error control.
More conservative choices, such as 1-SE rule $\lambda_{\mathrm{1SE}}$ or BIC-type criteria $\lambda_{\mathrm{BIC}}$, promote
sparser models and often stabilize error rates, but at the cost of reduced power and, in
some settings, slight undercoverage.
Importantly, these trade-offs were qualitatively stable across sample sizes and
censoring proportions, suggesting that they reflect systematic properties of the
methods rather than artifacts of specific simulation configurations. Notably, these pronounced differences in inferential behavior are only weakly reflected
in predictive performance, as IBS values remain comparatively stable across tuning
strategies.

Censoring further amplifies these differences.
As expected, increasing the censoring proportion reduces effective information and
decreases selective power across all methods.
At the same time, censoring magnifies method-specific sensitivities, particularly for
procedures that rely on strong conditioning or approximate variance estimation.
This observation underscores the importance of evaluating selective inference methods
under realistic censoring regimes, as commonly encountered in biomedical survival
studies.

The real-data analysis based on the METABRIC cohort illustrates these methodological
findings in a clinically relevant setting.
Rather than focusing on causal interpretation, the analysis highlights how selective
inference can be used to distinguish stable prognostic signals from weaker or unstable
associations.
Combining selection frequencies with selective confidence intervals provides a
transparent summary of both variable importance and uncertainty, and avoids overconfident
interpretation of effects that arise from data-adaptive modeling pipelines.
In this context, differences between inference methods are best understood as reflecting
different use cases: conservative approaches may be preferable when controlling false
discoveries is paramount, whereas more efficient procedures may be attractive in
exploratory settings where power and interval width are critical. Notably, the real-data analysis also suggests that the relative performance of selective inference methods may depend on covariate type, with differences observed between continuous and binary or ordinal predictors.

Our study has several limitations.
First, the simulation designs focus on a finite set of coefficient patterns and baseline
hazard specifications.
While we varied sample size, censoring, correlation structure, and tuning strategies,
more complex features such as time-varying effects, interactions, or non-proportional
hazards were not considered.
Second, our evaluation adopts a submodel-oriented view of inference.
Alternative estimands, such as full-model or target-population parameters under model
misspecification, may lead to different conclusions regarding the relative merits of the
methods.
Third, implementation details matter: default software choices were necessary for
comparability, but applied analyses may benefit from problem-specific calibration of
variance estimation, numerical tolerances, or tuning strategies.

Despite these limitations, the consistency of qualitative conclusions across a wide
range of scenarios provides reassurance that the main insights generalize to realistic
analysis workflows.
Overall, our results emphasize that selective inference can substantially improve the
credibility of post-selection conclusions in Cox models, but should be viewed as part of
a broader modeling strategy that includes careful tuning, sensitivity analyses, and
explicit communication of the intended inferential target.

\section*{Competing interests}

The authors declare no competing interests.

\section*{Funding}

Support by the DFG (grant FR 4121/2-2) is gratefully acknowledged.

\section*{Data availability}

All data analyzed in this study are publicly available from cBioPortal \cite{curtis2012genomic}.

\section*{Code availability}

The complete code used for the simulation study and data analysis is publicly available at
\url{https://github.com/lena-222/selective-inference-cox}.

\section*{Declaration of generative AI and AI-assisted technologies in the manuscript preparation process}
During the preparation of this work, the author(s) used ChatGPT solely for the purpose of improving the language and clarity of the manuscript. After using this tool, the author(s) reviewed and edited the content as needed and take(s) full responsibility for the content of the published article.

\section*{Author contributions}

L.S. conceived the study, implemented all methods, performed the analyses, and wrote the manuscript. S.F.-W. contributed to the methodological development and provided supervision, peer review, and critical feedback. All authors read and approved the final manuscript.

\bibliographystyle{unsrt}

\bibliography{references}

\begin{thebibliography}{10}

\bibitem{tibshirani1996regression}
Robert Tibshirani.
\newblock Regression shrinkage and selection via the {Lasso}.
\newblock {\em Journal of the Royal Statistical Society Series B: Statistical
  Methodology}, 58(1):267--288, 1996.

\bibitem{zou2006adaptive}
Hui Zou.
\newblock The {Adaptive Lasso} and its oracle properties.
\newblock {\em Journal of the American Statistical Association},
  101(476):1418--1429, 2006.

\bibitem{Berk2013PostSelection}
Richard Berk, Lawrence Brown, Andreas Buja, Kai Zhang, and Linda Zhao.
\newblock Valid post-selection inference.
\newblock {\em The Annals of Statistics}, 41(2):802--837, 2013.

\bibitem{lee2016exact}
Jason~D Lee, Dennis~L Sun, Yuekai Sun, and Jonathan~E Taylor.
\newblock Exact post-selection inference, with application to the {Lasso}.
\newblock {\em The Annals of Statistics}, 44(3):907--927, 2016.

\bibitem{andersen1993model}
Per~Kragh Andersen, {\O}rnulf Borgan, Richard~D Gill, and Niels Keiding.
\newblock Model specification and censoring.
\newblock In {\em Statistical Models Based on Counting Processes}, chapter~3,
  pages 121--175. Springer, New York, 1993.

\bibitem{tibshirani1997lasso}
Robert Tibshirani.
\newblock The {Lasso} method for variable selection in the {Cox} model.
\newblock {\em Statistics in Medicine}, 16(4):385--395, 1997.

\bibitem{zhang2007adaptive}
Hao~Helen Zhang and Wenbin Lu.
\newblock Adaptive {Lasso} for {Cox}'s proportional hazards model.
\newblock {\em Biometrika}, 94(3):691--703, 2007.

\bibitem{cox1975note}
David~R Cox.
\newblock A note on data-splitting for the evaluation of significance levels.
\newblock {\em Biometrika}, 62(2):441--444, 1975.

\bibitem{taylor2015statistical}
Jonathan Taylor and Robert~J Tibshirani.
\newblock Statistical learning and selective inference.
\newblock {\em Proceedings of the National Academy of Sciences},
  112(25):7629--7634, 2015.

\bibitem{yu2021coxci}
Yi~Yu, Jelena Bradic, and Richard~J Samworth.
\newblock Confidence intervals for high-dimensional {Cox} models.
\newblock {\em Statistica Sinica}, 31(1):243--267, 2021.

\bibitem{kong2021robustcox}
Shengchun Kong, Zhuqing Yu, Xianyang Zhang, and Guang Cheng.
\newblock High-dimensional robust inference for {Cox} regression models using
  desparsified {Lasso}.
\newblock {\em Scandinavian Journal of Statistics}, 48(3):1068--1095, 2021.

\bibitem{fithian2014optimal}
William Fithian, Dennis Sun, and Jonathan Taylor.
\newblock Optimal inference after model selection, 2017.

\bibitem{taylor2018post}
Jonathan Taylor and Robert Tibshirani.
\newblock Post-selection inference for $\ell_1$-penalized likelihood models.
\newblock {\em Canadian Journal of Statistics}, 46(1):41--61, 2018.

\bibitem{zhang2014confidence}
Cun-Hui Zhang and Stephanie~S Zhang.
\newblock Confidence intervals for low dimensional parameters in high
  dimensional linear models.
\newblock {\em Journal of the Royal Statistical Society Series B: Statistical
  Methodology}, 76(1):217--242, 2014.

\bibitem{vandegeer2014optimal}
Sara {van de Geer}, Peter B{\"u}hlmann, Ya'acov Ritov, and Ruben Dezeure.
\newblock On asymptotically optimal confidence regions and tests for
  high-dimensional models.
\newblock {\em The Annals of Statistics}, 42(3):1166--1202, 2014.

\bibitem{Kammer2022LassoSelective}
Michael Kammer, Daniela Dunkler, Stefan Michiels, and Georg Heinze.
\newblock Evaluating methods for {Lasso} selective inference in biomedical
  research: a comparative simulation study.
\newblock {\em BMC Medical Research Methodology}, 22(1):206, 2022.

\bibitem{cox1972regression}
David~R Cox.
\newblock Regression models and life-tables.
\newblock {\em Journal of the Royal Statistical Society: Series B
  (Methodological)}, 34(2):187--220, 1972.

\bibitem{fan2002variable}
Jianqing Fan and Runze Li.
\newblock Variable selection for {Cox}'s proportional hazards model and frailty
  model.
\newblock {\em The Annals of Statistics}, 30(1):74--99, 2002.

\bibitem{tang2017spikeslab}
Zaixiang Tang, Yueping Shen, Xinyan Zhang, and Nengjun Yi.
\newblock The spike-and-slab lasso cox model for survival prediction and
  associated genes detection.
\newblock {\em Bioinformatics}, 33(18):2799--2807, 2017.

\bibitem{wang2025lassocox}
Liwei Wang, Yu~Chang, Jinfeng Ma, Wenqing Qu, and Yifan Li.
\newblock Identifying high-risk candidates for prolonging progression-free
  survival in primary gastric carcinoma subject to ``double invasion'': an
  analytical approach utilizing lasso-cox regression.
\newblock {\em BMC Cancer}, 25(1):381, 2025.

\bibitem{Wainwright2009Sharp}
Martin~J. Wainwright.
\newblock Sharp thresholds for high-dimensional and noisy sparsity recovery
  using {$\ell_1$}-constrained quadratic programming (lasso).
\newblock {\em IEEE Transactions on Information Theory}, 55(5):2183--2202,
  2009.

\bibitem{hastie2015statistical}
Trevor Hastie, Robert Tibshirani, and Martin Wainwright.
\newblock {\em Statistical Learning with Sparsity: The Lasso and
  Generalizations}.
\newblock Chapman \& Hall/CRC, Boca Raton, FL, 2015.
\newblock Chapman \& Hall/CRC Monographs on Statistics and Applied Probability.

\bibitem{LeebPotscher2005}
Hannes Leeb and Benedikt~M. P{\"o}tscher.
\newblock Model selection and inference: Facts and fiction.
\newblock {\em Econometric Theory}, 21(1):21--59, 2005.

\bibitem{LeebPotscher2006}
Hannes Leeb and Benedikt~M. P{\"o}tscher.
\newblock Can one estimate the conditional distribution of post-model-selection
  estimators?
\newblock {\em The Annals of Statistics}, 34(5):2554--2591, 2006.

\bibitem{therneau2000cox}
Terry~M. Therneau and Patricia~M. Grambsch.
\newblock {\em Modeling Survival Data: Extending the {Cox} Model}.
\newblock Springer, New York, 2000.

\bibitem{tian2018selective}
Xiaoying Tian and Jonathan Taylor.
\newblock Selective inference with a randomized response.
\newblock {\em The Annals of Statistics}, 46(2):679--710, 2018.

\bibitem{xia2023coxdiverging}
Lu~Xia, Bin Nan, and Yi~Li.
\newblock Statistical inference for cox proportional hazards models with a
  diverging number of covariates.
\newblock {\em Scandinavian Journal of Statistics}, 50(2):550--571, 2023.

\bibitem{morris2019using}
Tim~P. Morris, Ian~R. White, and Michael~J. Crowther.
\newblock Using simulation studies to evaluate statistical methods.
\newblock {\em Statistics in Medicine}, 38(11):2074--2102, 2019.

\bibitem{ramos2020sampling}
P.~L. Ramos, D.~C.~F. Guzman, A.~L. Mota, D.~A. Saavedra, F.~A. Rodrigues, and
  F.~Louzada.
\newblock Sampling with censored data: a practical guide.
\newblock {\em Journal of Statistical Computation and Simulation},
  94(18):4072--4106, 2024.

\bibitem{andersen1982cox}
P.~K. Andersen and R.~D. Gill.
\newblock Cox's regression model for counting processes: A large sample study.
\newblock {\em Annals of Statistics}, 10(4):1100--1120, 1982.

\bibitem{curtis2012genomic}
Christina Curtis, Sohrab~P. Shah, Suet-Feung Chin, Gulisa Turashvili, Oscar~M.
  Rueda, Mark~J. Dunning, Doug Speed, Andy~G. Lynch, Shamith Samarajiwa, Yinyin
  Yuan, Stefan Gr{\"a}f, Gavin Ha, Gholamreza Haffari, Ali Bashashati, Roslin
  Russell, Steven McKinney, Anita Langer{\o}d, Andrew Green, Elena Provenzano,
  Gordon Wishart, Sarah Pinder, Peter Watson, Florian Markowetz, Leigh Murphy,
  Ian Ellis, Arnie Purushotham, Anne-Lise B{\o}rresen-Dale, James~D. Brenton,
  Simon Tavar{\'e}, Samuel Aparicio, and Carlos Caldas.
\newblock The genomic and transcriptomic architecture of 2,000 breast tumours
  reveals novel subgroups.
\newblock {\em Nature}, 486(7403):346--352, 2012.

\bibitem{sauer2025parametric}
Christina Sauer, F.~Julian~D. Lange, Maria Thurow, Ina Dormuth, and Anne-Laure
  Boulesteix.
\newblock Statistical parametric simulation studies based on real data, 2025.

\bibitem{tibshirani2017packageselectiveinference}
Ryan Tibshirani, Rob Tibshirani, Jonathan Taylor, Joshua Loftus, Stephen Reid,
  and Jelena Markovic.
\newblock {\em selectiveInference: Tools for Post-Selection Inference}, 2019.
\newblock R package version 1.2.5.

\bibitem{hastie2009elements}
Trevor Hastie, Robert Tibshirani, and Jerome Friedman.
\newblock {\em The Elements of Statistical Learning}.
\newblock Springer, New York, NY, USA, 2 edition, 2009.

\bibitem{akaike1974new}
Hirotugu Akaike.
\newblock A new look at the statistical model identification.
\newblock {\em IEEE Transactions on Automatic Control}, 19(6):716--723, 1974.

\bibitem{schwarz1978estimating}
Gideon Schwarz.
\newblock Estimating the dimension of a model.
\newblock {\em The Annals of Statistics}, 6(2):461--464, 1978.

\bibitem{fan2001variable}
Jianqing Fan and Runze Li.
\newblock Variable selection via nonconcave penalized likelihood and its oracle
  properties.
\newblock {\em Journal of the American Statistical Association},
  96(456):1348--1360, 2001.

\bibitem{Blanche2019Cindex}
Paul Blanche, Michael~W. Kattan, and Thomas~A. Gerds.
\newblock The c-index is not proper for the evaluation of t-year predicted
  risks.
\newblock {\em Biostatistics}, 20(2):347--357, 2019.

\bibitem{gerds2006consistent}
Thomas~A. Gerds and Martin Schumacher.
\newblock Consistent estimation of the expected brier score in general survival
  models with right-censored event times.
\newblock {\em Biometrical Journal}, 48(6):1029--1040, 2006.

\bibitem{therneau2013rpackagesurvival}
Terry~M Therneau.
\newblock {\em A Package for Survival Analysis in R}, 2024.
\newblock R package version 3.7-0.

\bibitem{friedman2021packageglmnet}
Jerome Friedman, Trevor Hastie, Robert Tibshirani, Balasubramanian Narasimhan,
  Kenneth Tay, Noah Simon, and Junyang Qian.
\newblock {\em glmnet: Lasso and Elastic-Net Regularized Generalized Linear
  Models}, 2021.
\newblock R package version 4.1-1.

\bibitem{glmnetold}
Jerome Friedman, Trevor Hastie, and Robert Tibshirani.
\newblock Regularization paths for generalized linear models via coordinate
  descent.
\newblock {\em Journal of Statistical Software}, 33(1):1--22, 2010.

\bibitem{coxnet}
Noah Simon, Jerome Friedman, Trevor Hastie, and Robert Tibshirani.
\newblock Regularization paths for cox's proportional hazards model via
  coordinate descent.
\newblock {\em Journal of Statistical Software}, 39(5):1--13, 2011.

\end{thebibliography}

\end{document}